\documentstyle[12pt]{article}
\def\lsim{\mathrel{\rlap {\raise.5ex\hbox{$ < $}}
{\lower.5ex\hbox{$\sim$}}}}

\newcommand{\pr}{\paragraph{}}
\newcommand{\be}{\begin{equation}}
\newcommand{\ee}{\end{equation}}
\newcommand{\bea}{\begin{eqnarray}}
\newcommand{\nn}{\nonumber}
\newcommand{\eea}{\end{eqnarray}}
\newcommand{\nd}[1]{/\hspace{-0.6em} #1}
\newcommand{\nk}{\noindent}
\def\gappeq{\mathrel{\rlap {\raise.5ex\hbox{$>$}}
{\lower.5ex\hbox{$\sim$}}}}

\def\lappeq{\mathrel{\rlap{\raise.5ex\hbox{$<$}}
{\lower.5ex\hbox{$\sim$}}}}

\begin{document}

\begin{titlepage}
\begin{flushright}
ACT-09/95 \\
%CERN-TH.127/95 \\
CTP-TAMU-24/95 \\
ENSLAPP-A-524/95 \\
hep-ph/9505401 \\
\end{flushright}

\begin{centering}
\vspace{.1in}
{\large {\bf
Non-Critical String Theory Formulation of Microtubule Dynamics
and Quantum Aspects of Brain Function
}} \\
\vspace{.2in}
{\bf N.E. Mavromatos$^{a,\diamond}$} and
{\bf D.V. Nanopoulos$^{b}$}
\vspace{.2in}
\begin{flushleft}
$^{a}$ Laboratoire de Physique Th\'eorique
ENSLAPP (URA 14-36 du CNRS, associ\'ee \`a l' E.N.S
de Lyon, et au LAPP (IN2P3-CNRS) d'Annecy-le-Vieux),
Chemin de Bellevue, BP 110, F-74941 Annecy-le-Vieux
Cedex, France. \\
$^{b}$ Texas A \& M University, College Station, TX 77843-4242, USA
and Astroparticle Physics Group, Houston
Advanced Research Center (HARC), The Mitchell Campus,
Woodlands, TX 77381, USA. \\
%$^{c}$ CERN, Theory Division, Geneva 23 Switzerland
\end{flushleft}
\vspace{.03in}

{\bf Abstract} \\
\vspace{.1in}
\end{centering}
{\small Microtubule (MT) networks, subneural
paracrystalline cytosceletal structures,
seem to play a fundamental role in the
neurons. We cast here the complicated
MT dynamics in the form of a $1+1$-dimensional
non-critical string theory, thus enabling
us to provide a consistent quantum
treatment of MTs, including enviromental
{\em friction} effects. We suggest, thus,
that the MTs are the microsites, in
the brain, for the emergence of stable, macroscopic
quantum coherent states,
identifiable with the {\em preconscious states}.
Quantum space-time effects, as
described by non-critical string theory, trigger
then an {\em organized collapse}
of the coherent states
down to a specific or {\em conscious state}. The whole
process we estimate to
take ${\cal O}(1\,{\rm sec})$, in excellent agreement
with a plethora of experimental/observational
findings. The {\em microscopic
arrow of time}, endemic
in non-critical string theory, and
apparent here in
the self-collapse process, provides
a satisfactory and simple resolution to
the age-old problem of how the, central to our feelings
of awareness, sensation
of the progression of time is generated.}

\vspace{0.1in}
\begin{flushleft}
ACT-09/95 \\
%CERN-TH.127/95 \\
CTP-TAMU-24/95 \\
ENSLAPP-A-524/95 \\
May 1995  \\
\end{flushleft}
\vspace{.01in}
\pr
\nk $^{\diamond}$ On leave from
P.P.A.R.C. Advanced Fellowship, Dept. of Physics
(Theoretical Physics), University of Oxford, 1 Keble Road,
Oxford OX1 3NP, U.K.  \\

\end{titlepage}
%\end{document}
\newpage
\section{Introduction}
\pr
The interior of living cells is structurally
and dynamically organized by {\it cytoskeletons}, i.e.
networks of protein polymers. Of these structures,
{\it MicroTubules} (MT) appear to be \cite{hameroff}
the most fundamental. Their dynamics has been
studied recently by a number of authors in connection
with the mechanism responsible for
dissipation-free
energy transfer. Recently,
Hameroff and Penrose \cite{HP} have conjectured another
fundamental r${\hat o}$le for the MT, namely being
responsible for {\it quantum computations}
in the human brain, and, thus, related to the
consciousness of the human mind.  The latter is argued to be
associated with certain aspects of quantum theory \cite{penrose}
that are believed to occur in the cytoskeleton MT, in particular
quantum superposition and subsequent collapse of the
wave function of coherent MT networks.
While quantum superposition is a well-established and well-understood
property of quantum physics, the collapse of the  wave function has been
always enigmatic. We propose here to use an explicit string-derived
mechanism -
in one {\it interpretation} of non-critical string theory -
for the collapse of the  wave function\cite{emn},
involving quantum
gravity in an
essential way and solidifying previous intuitively plausible
suggestions\cite{ehns,emohn}.
It is an amazing surprise that quantum gravity effects, of
order of magnitude $G_N^{1/2}m_{p}\sim10^{-19}$, with $G_N$ Newton's
gravitational constant and $m_{p}$ the proton mass,
can play a r\^ole in such low energies as the $eV$ scales
of the typical energy transfer that occurs in cytoskeleta.
However,
as we show in this article, the fine details of the MT characteristic
 structure
indicate that not only is this conceivable, but such
scenaria
lead to order of magnitude estimates for the
time scales entering {\it conscious perception}  that are close enough
to those conjectured/``observed'' by neuroscientists,
based on completely different grounds.
\pr
To understand how quantum space-time effects can affect
conscious perception, we mention that
it has long been suspected \cite{Frohlich}
that large scale quantum coherent phenomena can occur
in the interior of biological cells, as a result of the existence
of ordered water molecules. Quantum mechanical vibrations
of the latter are
responsible for the appearance of
`phonons' similar in nature to those associated with
superconductivity. In fact there is a close analogy between
superconductivity and energy transfer in biological cells.
In the former
phenomenon electric current is transferred without dissipation
in the surface of the superconductor. In biological cells, as we shall
discuss later on, energy is transferred through the cell
without {\it loss},
despite the existence of frictional forces that represent
the interaction of the cell with the surrounding water
molecules \cite{lal}.
Such large scale quantum coherent states can maintain
themselves for up to ${\cal O}(1\,{\rm sec})$,
%$500msec$
without significant
environmental entanglement. After that time, the state
undergoes self-collapse, probably
due to quantum
gravity effects. Due to quantum transitions
between the different states
of the quantum system
of MT in certain parts of the human brain,
a sufficient distortion of the surrounding space-time
occurs,
so that a microscopic (Planck size) black hole is formed.
Then collapse is induced, with a collapse time that
depends on the order of magnitude of the number $N$ of
coherent
microtubulins. It is estimated that, with an $N=O[10^{12}]$,
the collapse time
of
${\cal O}(1\,{\rm sec})$,
which appears to be
a typical time scale of conscious events, is achieved.
Taking into account that experiments have shown that
there exist
$N=10^{8}$ tubulins per neuron, and that there are $10^{11}$
neurons in the brain,
it follows that
that this order of magnitude for $N$ refers to a
fraction $10^{-7}$ of the human brain, which is very close to
the fraction believed responsible for human perception.
\pr
The self-collapse of the MT coherent state  wave function
is an essential step for the operation of the MT network
as a quantum computer. In the past it has been suggested
that MT networks processed information in a way similar
to classical cellular automata (CCA)
\cite{hamcel}. These
are described by interacting Ising spin chains on the spatial
plane obtained by fileting open and flattening the MT cylindrical
surface. Distortions in the configurations of individual
parts of the spin chain can be influenced by the environmental spins,
leading to information processing.
In view of the suggestion
\cite{HP} on viewing the conscious parts of the human mind
as quantum computers,
one might extend the concept of the
CCA to a quantum cellular automaton (QCA), undergoing
wave function self-collapses due to (quantum gravity)
enviromental entanglement.
\pr
An interesting and basic
issue that
arises in connection with the above r\^ole of the
brain as a quantum computer
is the emergence of a
direction in the flow of time (arrow). The
latter could be the result
of succesive self-collapses of the system's  wave function.
In a recent series of papers
\cite{emn} we have suggested a rather detailed
mechanism by which an {\it irreversible}
time variable has emerged in certain models of string  quantum
gravity. The model utilized string particles propagating
in singular space-time backgrounds with event horizons.
Consistency of the string approach requires conformal invariance
of the associated $\sigma$-model, which in turn implies a
coupling of the backgrounds for the propagating string modes
to an infinity of
global (quasi-topological) delocalized modes
at higher (massive)
string levels.
The existence of such couplings is necessitated by
specific coherence-preserving target space gauge symmetries
that mix the string levels \cite{emn}.
\pr
The specific model of ref. \cite{emn}
is a completely
integrable string theory, in the sense of being
characterised by an infinity of conserved charges.
This can be intuitively understood by the fact that
the model is a $(1+1)$-dimensional
Liouville string, and as such it can be mapped
to a theory of essentially free fermions
on a discretized world sheet
(matrix model approach \cite{matrix}).
A system of free fermions in $(1+1)$ dimensions
is trivially completely integrable, the infinity of
the conserved charges being provided by
appropriate moments of the fermion energies above the
Fermi surface. Of course, formally, the
precise symmetries of the model used in
ref. \cite{emn}
are much more
complicated \cite{bakas}, but the idea behind the
model's integrability is essentially the above.
It is our belief that
this quantum integrability is a very
important feature of theories of space-time
associated with the time arrow.
In its presence, theories with singular backgrounds
appear consistent as far as maintainance of
quantum coherence is concerned.
This is due to the fact that the phase-space density of the
field theory associated with the matter
degrees of freedom evolves with time according to the
conventional Liouville theorem\cite{emn}
\be
   \partial _t \rho = -\{\rho , H\}_{PB}
\label{one}
\ee
as a consequence of phase-space
volume-preserving symmetries.
In the two-dimensional example of ref. \cite{emn},
these symmetries are known as $W_{\infty}$, and are
associated
with higher spin target-space states\cite{bakas}. They are
responsible for string-level mixing, and hence they are
broken in any low-energy approximation.
If the concept of `measurement
by local scattering experiments' is introduced \cite{emn},
it becomes clear
that the observable
background cannot contain such global modes. The latter have to
be integrated out
in any effective
low-energy theory. The result of this integration
is a non-critical string theory,
based on the propagating modes only.
Its conformal invariance on the world sheet is  restored
by dressing the matter backgrounds by the
Liouville mode $\phi$, which plays the role of the time coordinate.
 The $\phi$ mode is a dynamical local
world-sheet scale \cite{emn},
flowing irreversibly as a result of certain theorems
of the renormalization group of unitary $\sigma$-models
\cite{zam}.
In this way time in target space
has a natural arrow for very specific {\it stringy reasons}.
\pr
Given the suggestion of ref. \cite{HP}
that space-time environmental entanglement could be
responsible for conscious brain function, it is natural to
examine the conditions under which our theory \cite{emn}
can be applied.
Our approach utilizes extra degrees of freedom, the $W_\infty$
global string modes, which are not directly
accessible to local scattering `experiments' that make
use of {\it propagating} modes only.
Such degrees of freedom carry information, in a similar spirit
to the information loss suggested by Hawking\cite{hawk}
for the quantum-black-hole case. For us, such degrees of freedom
are not exotic, as suggested in ref. \cite{Page},
but
appear {\it already} in the
non-critical String Universe \cite{emn,dn},
and as such they are considered as `purely stringy'.
In this respect, we believe that
the suggested model of consciousness, based on the
non-critical-string
formalism of ref. \cite{emn}, is physically more {\it concrete}.
The idea of using string theory instead of
point-like quantum gravity is primarily associated with the
fact that a {\it consistent} quantization of gravity
is at present possible {\it only} within the
framework of string theory, so far. However,
there are additional reasons that make advantageous
a string formalism. These include the possibility of
construction of
a completely integrable model for $MTs$, and
the Hamiltonian
representation of dynamical problems with friction involved in
the physics of MTs. This
leads to
the possibility of a consistent ({\it mean field}) {\it
quantization}
of certain soliton solutions associated with the energy
transfer mechanism in biologcal cells.
\pr
According to our previous discussion
emphasizing the importance of strings,
it is imperative that
we try to
identify the completely integrable
system underlying MT networks. Thus,
it appears essential to review
first the classical model for energy transfer
in biological systems associated with MT. This will allow
the identification of the
analogue of the (stringy)
propagating degrees of freedom, which
eventually couple to quantum (stringy) gravity
and to global environmental modes.
As we shall argue in subsequent sections, the relevant basic
building blocks of the human brain are one-dimensional
Ising spin chains,
interacting among themselves in a way so as to
create a large scale quantum coherent state,
believed to be responsible
for preconscious behaviour in the model of \cite{HP}.
The system can be described in a world-sheet
conformal invariant way and is unitary.
Coupling to gravity generates deviations from conformal
invariance which lead to time-dependence, by identifying
time with the Liouville field on the world sheet.
The situation
is similar to the environemtnal entanglement
of ref. \cite{vernon,cald}.
Due to this entanglement,
the system of the propagating modes opens up as in
Markov processes \cite{davidoff}.
This
leads to a dynamical self-collapse of the wave function
of the MT quantum coherent network.
In this way, the part of the human brain associated with
conscioussness generates, through successive collapses,
an arrow of time.
The
interaction among the spin chains,
then, provides a
mechanism for quantum computation,
resembling a planar
cellular automaton.
Such operations sustain the
irreversible flow of time.
\pr
The structure of the article is as follows:
in section 2 we discuss a model used for the
physical description of a MT, and in particular
for a
simulation of the energy transfer mechanism.
The model
can be expressed in terms of a $1+1$-dimensional
{\it classical field}, the projection of the
displacement field of
the MT dimers along the tubulin axis.
There exists {\it friction} due to intreraction
with the environment. However,
the theory possesses travelling-wave solitonic
states
responsible for loss-free transfer of energy.
In section 3, we give a formal representation
of the above system as a $1+1$-dimensional $c=1$
Liouville (string) theory. The advantage
of the method lies in that it allows for a
canonical quantization of the friction problem,
thereby yielding a model for a large-scale coherent state,
argued to simulate the preconscious state.
There is no time arrow in the above system.
In section 4, we discuss our mechanism
of introducing a dynamical time variable
with an arrow into the system, by elevating the
above $c=1$ Liouville theory to a $c=26$ non-critical
string theory, incorporating quantum
gravity effects. Such effects arise from the distortion
of space-time due to abrupt conformational
changes in the dimers. Such a coupling
leads
to a breakdown of the quantum coherence
of the preconscious state.
Estimates of the collapse times are given,
with the result that in this approach
concious perception of a time scale of
${\cal O}(1\,{\rm sec})$,
is due to a $10^{-7}$ part of the total brain.
In section 5 we briefly discuss
{\it growth} of a MT network in our framework,
which would be the analogue of
a non-critical string driven inflation
for the effective one-dimensional
universe of the MT dimer degrees of freedom.
We view MT {\it growth }
as an out-of-equilibrium one-dimensional
spontaneous-symmetry breaking process
and discuss the connection of
our approach to
some elementary theoretical
models with driven diffusion
that could serve as prototypes for
the phenomenon.
Conclusions and outlook are presented in section 6.
We discuss some technical apsects of our approach
in two
Appendices.

\section{Physical Description of the Microtubules}
\pr
In this section
we review certain features
of the MT that will be useful in subsequent parts
of this work. MT are hollow cylinders (cf Fig 1)
comprised of an exterior surface
(of cross-section
diameter
$25~nm$)
with 13 arrays
(protofilaments)
of protein
dimers
called tubulins.
The interior of the cylinder
(of cross-section
diameter $14~nm$)
contains ordered water molecules,
which implies the existence
of an electric dipole moment and an electric field.
The arrangement of the dimers is such that, if one ignores
their size,
they resemble
triangular lattices on the MT surface. Each dimer
consists of two hydrophobic protein pockets, and
has an electron.
There are two possible positions
of the electron, called $\alpha$ and $\beta$ {\it conformations},
which are depicted in Fig. 2. When the electron is
in the $\beta$-conformation there is a $29^o$ distortion
of the electric dipole moment as compared to the $\alpha $ conformation.
\pr
In standard models for the simulation of the MT dynamics,
the `physical' degree of freedom -
relevant for the description of the energy transfer -
is the projection of the electric dipole moment on the
longitudinal symmetry axis (x-axis) of the MT cylinder.
The $29^o$ distortion of the $\beta$-conformation
leads to a displacement $u_n$ along the $x$-axis,
which is thus the relevant physical degree of freedom.
This way, the effective system is one-dimensional (spatial),
and one has a first indication that quantum integrability
might appear. We shall argue  later on
that this is indeed the case.
\pr
Information processing
occurs via interactions among the MT protofilament chains.
The system may be considered as similar to a model of
interacting Ising chains on a trinagular lattice, the latter being
defined on the plane stemming from fileting open and flatening
the cylindrical surface of Fig. 1.
Classically, the various dimers can occur in either $\alpha$
or $\beta$ conformations. Each dimer is influenced by the neighboring
dimers resulting in the possibility of a transition. This is
the basis for classical information processing, which constitutes
the picture of a (classical) cellular automatum.
\pr
The quantum computer character of the MT network results
from the assumption that each dimer finds itself in a
superposition of $\alpha$ and $\beta$ conformations \cite{HP}.
There is a macroscopic
coherent state among the various chains, which
lasts for ${\cal O}(1\,{\rm sec})$ and constitutes the `preconscious'
state. The interaction
of the chains with (stringy) quantum gravity, then, induces
self-collapse of the wave function of the coherent MT network,
resulting in quantum computation.
\pr
In what follows we shall assume that the collapse occurs
mainly due to the interaction of each chain with
quantum gravity, the interaction from neighboring chains
being taken into account by including mean-field interaction terms
in the dynamics of the displacement field of each chain. This amounts
to a modification of the effective potential by
anharmonic oscillator terms.
Thus, the effective system under study is two-dimensional,
possesing one space and one time coordinate. The precise
meaning of `time' in our model will be clarified when
we discuss the `non-critrical string' representation
of our system.
\pr
Let $u_n$ be the displacement field of the $n$-th dimer in a MT
chain.
The continuous approximation proves sufficient for the study of
phenomena associated with energy transfer in biological cells,
and this implies that one can make the replacement
\be
  u_n \rightarrow u(x,t)
\label{three}
\ee
with $x$ a spatial coordinate along the longitudinal
symmetry axis of the MT. There is a time variable $t$
due to
fluctuations of the displacements $u(x)$ as a result of the
dipole oscillations in the dimers.
At this stage, $t$ is viewed as a reversible variable.
The effects of the neighboring
dimers (including neighboring chains)
can be phenomenologically accounted for by an effective
double-well potential \cite{mtmodel}
\be
U(u) = -\frac{1}{2}A u^2(x,t) + \frac{1}{4}Bu^4(x,t)
\label{four}
\ee
with $B > 0$. The parameter $A$ is temperature
dependent. The
model of ferroelectric distortive spin chains
of ref. \cite{collins} can be used to
give a temperature
dependence
\be
   A =-|cost|(T-T_c)
\label{temper}
\ee
where $T_c$ is a critical temperature of the system, and
the constant is determined phenomenologically \cite{mtmodel}.
In realistic cases the temperature $T$ is very close to $T_c$,
which for the human brain is taken to be the room temperature
$T_c = 300K$.
Thus, below $T_c$
$A > 0$.
The important relative minus sign in the potential (\ref{four}), then,
guarantees the necessary degeneracy, which is necessary for the
existence of
classical solitonic solutions. These constitute the basis
for our coherent-state description of the
preconscious state.
\pr
Including a phenomenological kinetic term for the dimers,
each having a mass $M$, one can write down a Hamiltonian
\cite{mtmodel}
\be
H = kR_0^2 (\partial _x u)^2  - M (\partial _t u)^2
-\frac{1}{2}A u^2 + \frac{1}{4}B u^4
+ qEu
\label{five}
\ee
where $k$ is a stiffness parameter, $R_0$ is the equilibrium
spacing between adjacent dimers, $E$ is the electric field
due to the `giant dipole' representation of the MT cylinder,
as suggested by the experimental results \cite{mtmodel},
and $q=18 \times 2e$ ($e$ the electron charge) is a
mobile charge. The spatial-derivative term in (\ref{five})
is a continuous approximation of terms in the lattice Hamiltonian
that express the effects of restoring strain forces between
adjacent dimers in the chains \cite{mtmodel}.
\pr
The effects of the surrounding water molecules can be
summarized by a viscuous force term that damps out the
dimer oscillations,
\be
 F=-\gamma \partial _t u
\label{six}
\ee
with $\gamma$ determined phenomenologically at this stage.
This friction should be viewed as an environmental effect, which
however does not lead to energy dissipation, as a result of the
non-trivial
solitonic structure of the
ground-state
and the non-zero constant
force due to the electric field.
This is a well known result, directly relevant to
energy transfer in biological systems \cite{lal}.
The modified equations of motion, then, read
\be
    M \frac{\partial ^2 u }{\partial t^2}
- k R_0^2 \frac{\partial ^2  u}{\partial x^2} - A u
+ B u^3 + \gamma \frac{\partial u}{\partial t} - qE =0
\label{seven}
\ee
According to ref. \cite{lal} the importance
of the force term $ qE $ lies in the fact
that eq (\ref{seven}) admits displaced classical soliton solutions
with no energy loss.
The solution acquires the form of a travelling wave,
and can
be most easily exhibited by defining a normalized
displacement field
\be
  \psi (\xi ) = \frac{u (\xi)}{\sqrt{A/B}}
\label{eight}
\ee
where,
\be
               \xi \equiv \alpha( x - v t)
\qquad \alpha \equiv \sqrt{\frac{|A|}{M(v_0^2 - v^2)^{\frac{1}{2}}}}
\label{nine}
\ee
with
\be
v_0 \equiv \sqrt{k/M} R_0
\label{sound}
\ee
the
sound velocity, of order $1 km/sec$,
and $v$ the propagation velocity
to be determined later.
In terms of the $\psi (\xi )$ variable,
equation (\ref{seven}) acquires the form of the equation
of motion of an anharmonic oscillator in a frictional environment
\bea
 \psi '' &+& \rho \psi ' - \psi ^3 + \psi + \sigma = 0  \nn \\
\rho &\equiv& \gamma v \sqrt{M |A|(v^2 - v_0^2)^{-\frac{1}{2}}}, \qquad
\sigma = q \sqrt{B}|A|^{-3/2}E
\label{ten}
\eea
which has a {\it unique} bounded solution \cite{mtmodel}
\be
    \psi (\xi ) = a + \frac{b -a}{1 + e^{\frac{b-a}{\sqrt{2}}\xi}}
\label{eleven}
\ee
with the parameters $b,a$ and $d$ satisfying:
\be
(\psi -a )(\psi -b )(\psi -d)=\psi ^3 - \psi -
\left(\frac{q \sqrt{B} }{|A|^{3/2}} E\right)
\label{twelve}
\ee
Thus, the kink propagates along the protofilament axis
with fixed velocity
\be
    v=v_0 [1 + \frac{2\gamma}{9d^2Mv_0^2}]^{-\frac{1}{2}}
\label{13}
\ee
This velocity depends on the strength of the electric
field $E$ through the dependence of $d$ on $E$ via (\ref{twelve}).
Notice that, due to friction, $v \ne v_0$, and this is essential
for a non-trivial second derivative term in (\ref{ten}), necessary
for wave propagation.
For realistic biological systems $v \simeq 2 m/sec$.
With a velocity of this order,
the travelling waves
of kink-like excitations of the displacment field
$u(\xi )$ transfer energy
along a moderately long microtubule
of length $L =10^{-6} m$ in about
\be
t_T = 5 \times 10^{-7} sec
\label{transfer}
\ee
This time is very close to Frohlich's time for
coherent phonons in biological system.
We shall come back to this issue later on.
\pr
The total energy of the solution (\ref{eleven}) is
easily calculated to be \cite{mtmodel}
\be
   E = \int _{-\infty}^{+\infty} dx H
= \frac{2\sqrt{2}}{3}\frac{A^2}{B} +
k \frac{A}{B} + \frac{1}{2} M^{*} v^2  \equiv \Delta + \frac{1}{2}
M^{*} v^2
\label{energy}
\ee
which is {\em conserved} in time.
The `effective' mass $M^{*}$ of the kink is given
by
\be
            M^{*} = \frac{4}{3\sqrt{2}}\frac{MA\alpha }{R_0 B}
\label{effmass}
\ee
The first term of equation (\ref{energy})
expresses the binding energy of the kink
and the second the resonant transfer energy.
In realistic
biological models the sum of these two terms
dominate over the third
term, being of order of $1eV$ \cite{mtmodel}. On the other hand,
the effective mass in (\ref{effmass}) is\cite{mtmodel} of order
$5 \times 10^{-27} kg$, which is about
the proton mass ($1 GeV$) (!).
As
we shall discuss later on, these values are essential
in yielding the correct estimates for the
time of collapse of the `{\it preconscious}' state due to our
quantum gravity environmental entangling. To make plausible
a consistent
study of such effects, we now discuss the possibility of
representing the equations of motions (\ref{ten})
as being derived from string theory.
\pr
Before closing we mention that the above {\it classical}
kink-like excitations (\ref{eleven})
have been discussed so far in connection
with physical mechanisms associated with
the
hydrolysis of GTP (Guanosine-ThreePhosphate)
tubulin dimers to GDP (Guanosine-DiPhosphate) ones.
Because the two forms of tubulins correspond
to different conformations $\alpha$ and $\beta$
above, it is conceivable to speculate that
the quantum mechanical oscillations between
these two forms of tubulin dimers
might be associated with a quantum version
of kink-like excitations in the MT network.
This is the idea we put forward in the present
work. The novelty of our approach
is the use of Liouville (non-critical) string theory
for the study of the dynamics involved.
This is discussed in the next section.

\section{Non-Critical (Liouville)
String Theory Representation of a MT}
\pr
It is important to notice that the relative sign (+)
between the second derivative and the linear term in $\psi $
in
equation (\ref{ten}) are such that this equation
can be considered as corresponding to the
tachyon $\beta$-function equation
of
a (1 + 1)-dimensional string theory, in a
flat space-time with a dilaton field $\Phi $
linear in the {\it space-like}
coordinate $\xi $ \cite{aben},
\be
        \Phi = - \rho \xi
\label{dilaton}
\ee
Indeed, the most general form of a `tachyon' deformation
in such a string theory, compatible with
conformal invariance is that of a travelling wave \cite{polch}
$T (x ' )$ , with
\bea
  x'=\gamma _v (x - vt) \qquad &;& \qquad t'= \gamma _v(t-vx) \nn \\
\gamma _v \equiv (1 &-& v^2)^{-1/2}
\label{st7}
\eea
where $v$ is the propagation velocity.
As argued in ref. \cite{polch} these translational invariant
configurations
are the most general
backgrounds,
consistent
with a {\it unique}
factorisation of the string $\sigma$-model
theory on a Minkowski space-time $G_{\mu\nu} = \eta _{\mu\nu}$
\be
  S= \frac{1}{4\pi \alpha '} \int d^2 z
  \left[\partial X^{\mu} {\overline \partial } X^{\nu}
G_{\mu\nu}(X) + \Phi(X) R^{(2)} + T(x)\right]
\label{st2b}
\ee
into two conformal field theories,
for the $t'$ and $x'$ fields,
corresponding to central charges
\be
c_{t'} = 1 -24 v^2 \gamma _v^2  \qquad ; \qquad
c_{x'} = 1 + 24 \gamma _v^2
\label{st8}
\ee
In our case (\ref{ten}), the r\^ole of the space-like
coordinate $x'$
is played by $\xi $, and the velocity $v$ is the velocity
of the kink. The velocity of light in this effective string
model is replaced by the sound velocity $v_0$ (\ref{sound}),
and the friction coefficient
$\rho $ is expressed in terms of the central charge deficit
(\ref{st8})
\be
  \rho = \sqrt{\frac{1}{6}(c(\xi) - 1)} = 2
 \gamma
\label{qdef}
\ee
The space-like `boosted' coordinate $\xi$, thus, plays the r\^ole
of space in this effective/Liouville mode string theory.
\pr
The important advantage of formulating the MT system as
a $c=1$ string theory, lies in the possibility
of casting the friction problem in a Hamiltonian
form. To this end, we now make some comments on
the
various non-derivative terms in the
tachyon potential $V(T)\equiv U ( \psi ) $ in
the target-space effective action
\cite{banks}.
Such terms contributre higher order non-derivative terms
in the equations of motion (\ref{ten}). The term linear in $\psi $
is fixed in string theory and normalized (with respect to the
second derivative term ) as in (\ref{ten}) \cite{polch}.
The higher order  terms are polynomials in $\psi $ and
their coefficients can be varied according to
renormalization prescription \cite{banks}.
\pr
The general structure of the tachyon effective action in
the target space of the string is, therefore,
of the form \cite{banks}
\be
{\cal L} = e^{-\Phi }\sqrt{G} \left[ (T - c)^2 +
{\hat {\cal L}}(\nabla T , \nabla \Phi )\right]
\label{effact}
\ee
where $G _{\mu\nu}$ is a target space metric field,
and $c$ is a constant.
The linear term is the only universal term, showing
the impossibility of finding a stable tachyon background
in bosonic string theory, unless it is time dependent.
\pr
In our specific model, we have seeen that a term cubic in $\psi $
in the equation of motion (\ref{ten}), with a relative minus sign as compared
to
the linear term, was responsible for the appearance of a kink-like classical
 solution.
Any change in the non-linear terms would obviously affect the
structure of the solution, and we should understand
the physical meaning of this in our biological system.
To this end, we consider a general polynomial in $T$
equation of motion for a static tachyon
in (1+1) string theory
\be
  T '' (\xi ) + 2\gamma T' (\xi ) = P( T )
\label{polyn}
\ee
where $\xi $ is some space-like co-ordinate
and $P (T)$ is a polynomial of degree $n$, say.
The `friction' term $T '$ expresses
a Liouville derivative, since the effective
string theory of the displacement field $u$ is viewed as
a $c=1$
matter non-critical string. In our interpretation
of the Liouville field as a local scale on the world sheet
it is natural to assume that the single-derivative term
expresses the non-critical string $\beta$-function, and hence
is itself a polynomial  $R$ of degree $m$
\be
     T ' (\xi ) = R (T)
\label{secondpol}
\ee
Using Wilson's exact renormalization group scheme,
we may assume that $R(T) = a_2 T + a_4 T^2 $,
where  $a_2, a_4$ are related to the anomalous dimension
and operator product expansion coefficients for the tachyon
couplings.
The compatibility conditions for the existence of bounded
solutions to the equation (\ref{polyn}), then,
imply the form\cite{curiouseq}
$P(T) = A_1 + A_2 T + A_3 T^2 + A_4 T^4 $,
with $A_i = f_i (a_2, a_3), i =1,...4$.
Indeed such equations have been shown \cite{curiouseq}
to lead to kink-like solutions,
\be
  T (\xi ) = \frac{1}{2a_4} \{ {\rm sgn}(a_2a_4) a_2
{\rm tanh}[\frac{1}{2}a_2  (x-u t)]- a_2 \} \
\label{kink}
\ee
where the velocity $u=\frac{A_3 - 3a_2a_4}{a_4}$.
This fact
expresses for us
a sort of universal behaviour
for biological systems. This shows the existence of {\em at least}
one class of schemes which admit kink like solutions
of the same sort as the ones of Lal \cite{lal}
for energy transfer without dissipation in cells.
\pr
The importance of solutions of the form (\ref{kink}) lies in the
fact that they can be derived from a Hamiltonian and, thus, can be
quantized canonically \cite{hojman}.
They are connected to the solitons (\ref{eleven})
by a Renormalization Scheme change on the world-sheet of the
effective string theory, reproducing (\ref{ten}).
This
amounts to
the possibility of casting friction problems, due to the Liouville
terms, into a Hamiltonian form.
This is quite important for the quantization of the kink solution,
which will provide one with a concrete example of
a large-scale quantum coherent state for the preconscious
state of the mind.
In a pure field-theoretic setting, a quantization scheme
has been discussed in ref. \cite{tdva}, using
a variational approach by means of squeezed coherent
states. There is a vast literature on soliton
quantization using approximate $WKB$ methods.
We selected this method for our purposes here,
because it yields more
accurate results than the usual $WKB$ methods
of soliton quantization\cite{WKB}, and it seems
more appropriate for our purposes here, due to its
direct link with coherent ground states.
We shall not give details on the derivation but
concentrate on the results. We refer the
interested reader to the literature\cite{tdva}.
A brief description of the method
is provided in Appendix B.
\pr
The result of such a quantization was a
modified soliton equation for the (quantum corrected) field
$C(x,t)$ \cite{tdva}
\be
    \partial ^2 _t C(x.t) - \partial _x ^2 C(x,t)
+ {\cal M}^{(1)} [C(x,t)] = 0
\label{22c}
\ee
with the notation
\be
M^{(n)} = e^{\frac{1}{2}(G(x,x,t)-G_0(x,x))\frac{\partial ^2}
{\partial z^2}} U^{(n)}(z) |_{z=C(x,t)}
\qquad ; \qquad U^{(n)} \equiv d^n U/d z^n
\label{22}
\ee
Above, $U $ denotes the potential of the original soliton
Hamiltonian,
and $G (x,y,t)$ is a bilocal field that describes quantum corrections
due to the modified
boson field around the soliton.
The quantities $M^{(n)}$ carry information about the
quantum corrections, and in this sense the above scheme
is more accurate than
the WKB approximation \cite{tdva}.
The whole scheme may be thought of as a
mean-field-approach to quantum corrections to the soliton
solutions.
For the kink soliton (\ref{kink}) the quantum
corrections (\ref{22c})
have been calculated explicitly in ref. \cite{tdva},
thereby providing us with a  concrete
example of a large-scale quantum coherent state.
\pr
The above results on a consistent quantization of the soliton
solutions (\ref{kink}), derived from a Hamiltonian
function, find a much more general and simpler application
in our Liouville approach \cite{emninfl,emnest}.
To this end, we first note that
the structure of the equation
(\ref{polyn}), which leads to (\ref{kink}),
is generic for Liouville strings, with
arbitrary targets.
If we view the Liouville mode as a local scale
of the renormalization group on the world-sheet \cite{emn,osborn},
one can easily show that for the coupling $g^i$ of any
non-marginal deformation $V_i$ of the $\sigma$-model
\footnote{For a concise review of the formalism and relevant notation
see Appendix A.},
the following
Liouville renormalization group equation holds
\cite{tseytl,emn}
\be
  {\ddot g}^i + Q {\dot g}^i = -\beta ^i = - G^{ij}
  \partial _i C [g]
\qquad ; \qquad Q^2 = \frac{C[g]-25}{6} + \dots
\label{st5}
\ee
where the dot denotes differentiation with respect to the
renormalization group Liouville scale $t$, and
$\dots $ denote terms removable by redefinitions
of the couplings $g^i$ (renormalization scheme changes).
The tensor $G^{ij}$ is an inverse metric in field space\cite{zam}.
Notice in (\ref{st5})
the r\^ole of the non-criticality of the string
($Q \ne 0$)
as providing a source of friction \cite{emn}
in the space of fields $g^i$.
The non-vanishing renormalization group $\beta$-functions
play the r\^ole of generalized forces.
The functional
$C[g]$ is the Zamolodchikov $c$-function, which is
constucted out of a particular combination
of components of the world-sheet stress  tensor
of the deformed $\sigma$-model \cite{zam,emn}.
\pr
The existence of friction terms in (\ref{st5})
implies a statistical description of the
temporal evolution of the system using
classical density matrices $\rho (g^i, t)$ \cite{emn}
\be
  \partial _t \rho  = -\{ \rho , H \}_{PB} +
\beta ^i G_{ij} \frac{\partial \rho }{\partial p_j}
\label{liouvmod}
\ee
where $p_i$ are conjugate momenta to $g^i$, and $G_{ij}\propto
<V_iV_j>$
is a metric in the space of fields $\{ g^i \}$ \cite{zam}.
The non-Hamiltonian term in (\ref{liouvmod})
leads to a violation of the Liouville theorem (\ref{one}) in the
classical phase  space $\{g^i, p_j \}$,
and constitutes the basis for a modified (dissipative)
quantum-mechanical
description of the system\cite{emn}, upon quantization.
\pr
In string theory, summation over world sheet surfaces
will imply quantum fluctuations of the string target-space
background fields $ g^i$\cite{emn,emninfl}.
{\it Canonical} quantization
in the space $\{ g^i \}$ can be achieved, given that
the necessary Helmholtz conditions \cite{hojman} can be shown to
hold in the string case \cite{emninfl}. The important feature
of the string-loop corrected (quantum)
conformal invariance conditions is that they can
be derived from a target-space action \cite{mm2},
which schematically can be represented as\cite{hojman,emninfl}
\bea
    {\cal S}  &=& -
    \int dt (\int _{0}^1 d\tau g^i E_i (t, \tau g, \tau {\dot g},
\tau {\ddot g} ) + total~derivatives \qquad ; \nn \\
 E_i (t,g,{\dot g},{\ddot g}) &\equiv &
 {\cal G}_{ij}( {\ddot g}^j
 + Q{\dot g}^i + \beta ^i )
\label{st6b}
\eea
where the tensor ${\cal G}_{ij}$ is a
(quantum) metric\cite{zam}
in theory space $\{ g ^i \}$. It
is characterized  by a specific behaviour \cite{emnest}
under the action of the renormalization group
operator ${\cal D } \equiv \partial _t + {\dot g}^i \partial _i$,
 \be
 {\cal D} {\cal G}_{ij} = Q {\cal G}_{ij} = <V_i V_j > + \dots
 \label{st6c}
\ee
where the $\dots $ denote diffeomorphism terms in $g$-space, that
can be removed
by an appropriate scheme choice\cite{emnest}.
\pr
In this way, a friction problem (\ref{st5})
can be mapped non-trivially onto a canonically
quantized Hamiltonian system, in similar spirit
to the solitonic point-like field theory
discussed in section 3.
The quantum version of (\ref{liouvmod})
reads \cite{emn}
\be
    \partial _t {\hat \rho} =
i [ {\hat \rho}, {\hat H} ] + i \beta ^i G_{ij} [ {\hat g}^i ,
{\hat \rho }]
\label{quantumliouv}
\ee
where the hat notation denotes quantum
operators, and appropriate quantum ordering is understood
(see below). We note that
the equation (\ref{quantumliouv})
implies that $\partial _t \rho$ dependes only on $\rho (t)$
and not on a particular decomposition
of $\rho (t)$  in the projections corresponding to various
results of a measurement process. This automatically
implies the absence of faster-then-light signals
during the evolution.
\pr
The analogy with the soliton case, discussed previously,
goes even
further if we recall \cite{emn} the fact that
energy is conserved in average
in our approach of time as a renormalization
scale.
Indeed, it can be shown that
the temporal change of the
energy functional of the string particle,
${\cal H}$, obeys the equation
\be
  \partial _t {\cal H} \propto {\cal D}<\Theta (z,{\overline z}),
\Theta (0) > = 0
\label{st6d}
\ee
where $\Theta $ denotes the trace of the world-sheet
stress tensor; the vanishing result is due to the
{\it renormalizability} of the $\sigma$-model on the
world-sheet surface, which, thus, replaces time-translation
invariance in target space.
\pr
However, the quantum energy fluctuations
$\delta E~\equiv~[<<{\cal H}^2>> - (<<{\cal H}>>)^2 ]^{\frac{1}{2}}$
are time-dependent :
\be
\partial _t (\delta E)^2
= -i <<[\beta ^i , {\cal H}]\beta^j G_{ji}>> =
<<\beta^j G_{ji} \frac{d}{d t}\beta ^i >>
\label{40}
\ee
Using the fact that $\beta ^i G_{ij} \beta ^j $ is a
renormalization-group invariant quantity,
we can express (\ref{40}) in the form
\be
  \partial _t (\delta E)^2  = -
<<Q^2  \beta ^i G_{ij} \beta ^j >>= -
<<Q^2  \partial _t C >> \le 0
\label{fluct}
\ee
We, thus, observe that,
for $Q^2 > 0$ (supercritical strings), the
energy fluctuations decrease
with time for unitary string
$\sigma$-models \cite{zam}.
\pr
Before closing this section we wish to make
an important remark concerning
quantum ordering in (\ref{quantumliouv}).
The quantum ordering is chosen in such a way  so
that energy and probability conservation,
and positivity of the density matrix
are preserved in the quantum case. Taking into account
that in our string case
$\beta ^j G_{ij} = \sum _{n} C_{ij_1\dots j_n}g^{j_1} \dots g^{j_n} $,
with the expansion coefficients appropriate vertex operator
correlation functions \cite{emn},
it is straightforward to cast the
the above equation into a Lindblad form~\cite{lin}
\be
{\dot \rho } \equiv \partial _t \rho
= i [\rho, H] - \{ B^\dagger B, \rho _t \}_{+}
+ 2  B\rho _t B^\dagger
\label{bloch}
\ee
where the `environment' operators $B$, $B^\dagger$
are defined appropriately as `squared roots'
of the various partitions of the operator
$\beta ^j G_{ij} \dots g^i $.
This form may have important consequences
in the case one considers a wave-function
representation of the density matrix.
Indeed, as discussed in ref. \cite{gisin},
(\ref{bloch}) implies a {\it stochastic} diffusion
equation for a state vector, which has important
consequences for the localization of the wave-function
in a quantum
theory of measurement. We stress, however,
that our approach based on density matrices
(\ref{bloch}) and renormalizability of the
string $\sigma$-model is more general
than any approach assuming state vectors.

\pr
\section{Quantum Gravity and
Breakdown of Coherence in the
String Picture of a Microtubule}
\pr
Above we have established the conditions
under which a large-scale coherent state
appears in the MT network, which can be
considered responsible for loss-free energy
transfer along the tubulins.
As a result
of conformational (quantum) transitions
of the tubulin
dimers, there is an
abrupt distortion of space-time.
Formally this is expressed by coupling the
$c=1$ string theory to two-dimensional quantum gravity.
This elevates the matter-gravity system to a
critical $c=26$ theory.
Such a coupling, then,
causes decoherence, due to induced instabilities
of the kink quantum-coherent `preconscious state',
in a way that we shall discuss below.
As
the required
collapse time of ${\cal O}(1\,{\rm sec})$ of
the wave function of the coherent
MT network is several orders of magnitude bigger than the
energy transfer
time  $t_T$ (\ref{transfer}),
the two mechanisms are compatible with each other.
Energy is transfered during the quantum-coherent
preconscious state, in $10^{-7} sec$,
and then collapse occurs to a certain (classical) conformational
configuration. In this way, Frohlich's frequency
associated with coherent `phonons' in biological
cells is recovered, but in a rather different setting.
\pr
We now proceed to describe the precise
mechanism for the breakdown of coherence,
once the system couples to quantum gravity.
First, we discuss an explicit way of
dynamical creation of black holes in two-dimensional
string theory \cite{russo,witt}
through
collapse of tachyonic matter $T(x,t)$.
This is a procedure that can happen
as a result of quantum fluctuations
of various excitations.
Consider the equations of motion
for the graviton and dilaton fields obtained by
imposing conformal invariance in the model
(\ref{st2b}) to order $\alpha '$, ignoring the non-universal
tachyon terms in the potential.
It can be shown that a generic solution for the graviton
deformation
has the form \cite{russo}
\be
 ds^2 = -\{1 + \int ^x _{\infty} dx'[(\partial _{x'} T)^2
+ ({\dot T})^2 ] - \int ^t _{-\infty} dt' {\dot T} \partial _{x'} T \}
dt^2 + \{1 + \int ^t _{-\infty} dt' {\dot T}\partial _{x'} T \}dx^2
\label{st9}
\ee
Consider now an incoming localized wavepacket of the form
($a \equiv {\rm const}$)
\be
T = e^{-x} \frac{a}{cosh [2(x + t)]}
\label{st10}
\ee
It becomes clear from (\ref{st9}) that there will be
an horizon, obtained as a solution of the equation
derived
by imposing
the vanishing
of the coefficient of the $dt^2 $ term . Thus,
for late times $ t \rightarrow \infty$, the resulting
metric configuration
is a two-dimensional static black hole \cite{witt}
\be
      ds^2 = -(1 - \frac{4}{3} a^2 e^{-2x} ) dt^2
+ (1 - \frac{4}{3}a^2 e^{-2x} )^{-1} dx^2
\label{st11}
\ee
and the whole process describes a dynamical collapse of
matter.
The energy of the collapsing
wavepacket gives rise to the
ADM mass $\frac{4}{3}a^2 $ of the black hole \cite{witt}.
It is crucial for the argument that there is no part
of the wave-packet reflected. Otherwise the resulting
ADM mass of the black hole will be the part of the
energy that was not reflected.
In realistic situations, the black hole is only {\it virtual},
since low-energy matter pulses are always reflected in two-dimensional
string theories, as suggested
by matrix-model
computations \cite{polchmatr}.
Indeed, if one discusses
pulses which undergo total reflection
\cite{russo},
\be
T (x,t) = e^{-x} \eta (x, t) \qquad ; \qquad
   \eta (x,t) =
   \frac{a}{cosh(2 (x + t))} +
   \frac{a}{cosh(2 (x - t))}
\label{refl}
\ee
then, it can be easily shown that there is
a {\it transitory} period where the space time geometry looks like a
black hole (\ref{st11}), but asymptotically in time
one recovers the linear dilaton (flat) vacuum\cite{aben}.
The
above example (\ref{refl})
gives a generic way of a (virtual)
dynamical matter
collapse in a two-dimensional stringy space-time,
of the type that we encounter in our model for the
brain, as a result
of quantum conformational changes of the
dimers.
In the case of MT,
the replacement $x \rightarrow \xi $, where $\xi $ is
the boosted coordinate in (\ref{eleven}),
is understood.
\pr
The important point to notice is that the
system of $T(\xi )$ coupled to a black hole
space-time (\ref{st11}),
even if the latter is a virtual configuration, it
cannot be critical
(conformal invariant) {\it non-perturbatively}
if the tachyon has a
travelling wave form. The factorisation
property of the world-sheet action  (\ref{st2b})
in the flat space-time case breaks down due to the
non-trivial graviton structure (\ref{st11}).
Then a travelling wave cannot be compatible with
conformal invariance, and renormalization scale dressing
appears necessary.
The gravity-matter system is viewed as a $c=26$ string
\cite{witt,emn}, and hence the
renormalization scale is {\it time-like} \cite{aben}.
This implies time dependence in $T (\xi, t)$.
\pr
A natural question arises whether there exist a deformation
that turns on the coupling $T (\xi )$ which is exactly
marginal so as to maintain conformal invariance.
To answer this question, we first note that there is
an exact conformal model \cite{witt},
a Wess-Zumino $SL(2,R)/U(1)$ coset theory, whose target
space has the metric (\ref{st11}). The exaclty marginal
deformation of this black hole background
that turns on matter, couples necessarily
the propagating tachyon $T(\xi )$ zero modes to
an infinity of higher-level string states \cite{chaudh}.
The latter are classified according to
discrete representations
of the $SL(2,R)$ isospin, and together with the
propagating modes,
form
a target-space $W_\infty$-algebra \cite{emn,bakas}.
This coupling of massive and massless modes
is due to the non-vanishing Operator Product Expansion
(O.P.E.)
among the vertex operators of the
$SL(2,R)/U(1)$ theory \cite{chaudh}. The
model is completely integrable,
due to an infinity of conserved charges in target space
\cite{emn} corresponding to the Cartan subalgebra
of the infinite-dimensional $W_\infty$ \cite{bakas}.
This integrability persists quantization \cite{wu},
and it is very important for the quantum coherence
of the string black hole space-time\cite{emn}. Due to the
specific nature of the $W_\infty$ symmetries, there
is no information loss during a stringy black hole decay,
the latter being
associated
with
instabilities
induced by higher-genus effects on the world-sheet
\cite{emndec,emn}. The phase-space volume of the
effective field theory is preserved in time, {\it only
if} the infinite set of the global string modes
is taken into account. This is due to the string-level mixing
property of the $W_\infty$ - symmetries of the target space.
\pr
However,
any local operation of measurement, based on local scattering
of propagating matter, such as the functions performed
by the human brain, will necessarily break this coherence,
due to the truncation of the string deformation spectrum
to the localized propagating modes $T(\xi)$. The latter will, then,
constitute a subsystem
in interaction with an
{\it environment} of global string modes.
The quantum integrability of the full string system is crucial in
providing the necessary couplings.
This breaking of coherence
results in an arrow of time/Liouville scale,
in the way
explained briefly
above \cite{emn}. The black-hole $\sigma$-model
is viewed as a $c=26$ critical string, while the travelling
wave background is a non-conformal deformation.
To restore criticality one has to
dress $T(\xi )$ with a Liouville time dependence
$T(\xi, t)$ \cite{emn}.
 From a $\sigma$-model point of view,
to $O(\alpha ')$, a non-trivial consistency check
of this approach
for the black hole model of ref. \cite{witt} has been
provided in ref. \cite{emn}.
We stress once more that
the Liouville renormalization scale now is time-like,
in contrast with the previous string picture
of a $c=1$ matter string theory, representing the
displacement field $u$ alone before coupling to gravity.
\pr
By viewing the time $t$ as a local scale on
the world sheet, a natural identification
of $t$
will be with the logarithm of the area $ A $ of
the world sheet, at a fixed topology.
As the non-critical string runs towards the
infrared fixed point the area expands.
In our approach to
Liouville time
\cite{emn}, the actual flow of time
is opposite to the world sheet renormalization
group flow. This is
favoured by a bounce interpretation of the
Liouville flow due to specific
regularization properties of Liouville correlation
functions \cite{emn,kogan}.
This imlplies that we may set $t \propto -ln A$, with $A$
flowing always towards the infrared $A \rightarrow \infty$.
In this way, a {\bf Time Arrow} is implemented
automatically in our approach, without requiring the
imposition of time-asymmetric boundary conditions in
the analogue of the Hartle-Hawking state. In this respect,
our theory has many similarities to models of
conventional dissipative
systems, whose mathematical formalism \cite{dissip,santilli}
finds a natural application to our case.
\pr
With this in mind, one can examine the
properties of the
correlation functions of $V_i$, $A_N = <V_{i_1} \dots V_{i_N} >$,
and hence the issue of coherence breakdown.
In critical string theory such correlators
correspond to scattering amplitudes in the target-space
theory. It is, therefore, essential to check on
this interpretation in the present situation.
Since the correlators are mathematically
formulated on fixed area $A$
world-sheets, through the so-called fixed area constraint
formalism \cite{DDK,mm}, it is interesting to look for
possible $A$-dependences in their evolution. In such a
case their interpretation as target-space scattering amplitudes
would fail.
Indeed, it has been shown \cite{emnest}
that there is an induced target
space $A$-dependence of the regularized correlator $A_N$,
which, therefore, cannot be identified with a
target space $S$-matrix element, as was the
case of critical strings \cite{stringbook}.
Instead, one has non-factorisable contributions
to a superscattering amplitude $\nd{S} \ne S S^{\dagger} $,
as is usually the case in {\it open} quantum mechanical systems,
where the fundamental building blocks are density matrices
and not pure quantum states\cite{ehns}.
For completeness, we describe in Appendix A
some formal aspects of this situation,
based on results of our approach to non-critical
strings \cite{emn}.
\pr
It is this sort of coherence breakdown that we advocate as
happening
inside the part of the brain related to {\it consciousness},
whose operation is
described by the dynamics of (the quantum version of) the model
(\ref{five}).
The effective two-dimensional substructures, that we have
identified above as the basic elements
for the energy transfer in MT, provide the necessary
framework for coupling the
(integrable) stringy
black hole space-time (\ref{st11})
to the displacement field $u(x,t)$. This allows for
a qualitative description of
the effects of
quantum gravity on the
coherent superposition of the preconcious states.
\pr
One can calculate in this approach
the off-diagonal
elements of the density matrix in the string theory space $u^i$,
with now $u^i(t)$ representing the displacement field of the
$i$-th dimer. In a $\sigma$-model representation  (\ref{st2b}),
this is the tachyon deformation.
The computation proceeds analogously \cite{emn}
to the
Feynman-Vernon \cite{vernon}  and
Caldeira and Leggett \cite{cald}
model of environmental oscillators, using the influence functional
method, generalized properly to the string theory
space\footnote{Throughout this work, as well as
our previous works on the subject \cite{emn}, we assume that
such a space exists, and admits \cite{zam} a metric
${\cal G}_{ij}$. Indications that this is true are obtained
from perturbative calculations in  $\sigma$-model, to which
we base our belief.}
$u_i$. The
general theory of time as a world-sheet
scale predicts \cite{emn}
the following
expression for the reduced density
matrix\cite{vernon,cald} of the observable states:
\bea
\nonumber     \rho (u_I,u_F,  t  ) / \rho_S (u_I,u_F,  t  ) \simeq
e^{ - N \int _0^{t} d\tau \int_{\tau-\epsilon}^{\tau + \epsilon}
d\tau '
\frac{(u(\tau) - u(\tau '))^2}{(\tau - \tau ')^2}} \simeq \\
e^{- N \int_0^{t} d\tau \int_{\tau ' \simeq \tau }
d\tau '
 {\hat \beta} ^i G_{ij}(S_0){\hat  \beta} ^j }
 \simeq
e^{ - D N t ({\bf u_I}  - {\bf  u_F} )^2 + \dots }
\label{asym}
\eea
\nk where
the subscript ``S'' denotes quantities evaluated
in conventional
Schr\"odinger
quantum mechanics, and $N$ is the
number of the environment `atoms'\cite{emohn}
interacting with the background $u^i$.
For $(1,1)$ operators, that we are interested in,
the structure of the
renormalization group ${\hat \beta}$-functions
is
${\hat \beta}^i=\epsilon u^i + \beta^{ui} $,
where $\beta^{ui}=O(u^2)$ and
$\epsilon \rightarrow 0$ is the anomalous dimension.
Recalling \cite{emn}
the pole structure of the
Zamolodchikov metric,
$
G_{ij}=\frac{1}{\epsilon}{\cal G}_{ij}^{(1)} + regular$,
one finds that the dominant contribution
to the exponent
$K$
of the model (\ref{asym})
comes
from the
$\epsilon$-term in ${\hat \beta}^i $ and the
pole term in $G_{ij}$ :
\be
K=N \int _0^t {\hat \beta}^i G_{ij} {\hat \beta}^j d\tau \ni
2N\int _0^t u^i{\cal G}_{ij}^{(1)}\beta^{uj} d\tau + O(\epsilon )
\label{dcoeff}
\ee
\pr
Assuming slowly varying $u^i$ and $\beta ^i$
over the time $t$,
this implies
that the off-diagonal
elements of the density matrix
would decay exponentially to zero, within a collapse
time of order \cite{emohn}
\be
t_{coll} =\frac{1}{N} (O[ \beta ^{u^i} {\cal G}_{ij}^{(1)} u^j])^{-1}
t_s
\label{collapse}
\ee
in fundamental string units $t_s$ of time. The
superscript $(1)$ in the theory space metric
denotes the single residue in, say, dimensional
regularization on the world sheet \cite{emn,osborn};
$N$ is the number
of (coherent) tubulin dimers in interaction with the given
dimer that undergoes the abrupt conformational change.
Here the $\beta ^{u^i}$ -function is assumed to admit
a perturbative
expansion in powers of $\lambda _{s} ^2 \partial _{X}^2 $,
in target space, where the fundamental string unit of length
is defined as
\be
 \lambda _{s} =(\frac{\hbar \alpha '}{v_0^2})^{\frac{1}{2}}
\label{length}
\ee
where $v_0$ is the sound velocity (\ref{sound}), of order
$1 km/sec$ \cite{mtmodel}.
We work
in a system of
units where the light velocity is
$c=1$,
and
we use as the scale where
quantum gravity effects become important,
the grand unified string scale,
$M_{gus} = 10^{18} GeV$,
or in length $10^{-32} cm$,
which is $10$ times the conventional
Planck scale.
This is so, because our
model is supposed to be an effective description of quantum gravity
effects in a stringy (and not point-like)
space-time.
This scale corresponds to a time scale of $t_{gus}
= 10^{-42} sec$.
We now observe that, to leading order in the perturbative
$\beta$ function expansion in (\ref{collapse}), any
dependence on the velocity $v^2$ disappears
in favour of the scales $M_{gus}$ and $t_{gus}$.
It is, then, straightforward to
obtain a rough estimate for the collapse time
\be
    t_{col} =O[\frac{M_{gus}}{E^2N}]
\label{colltime}
\ee
where $E$ is  a typical energy scale in the problem.
Thus, we
estimate
that a collapse time
of ${\cal O}(1\,{\rm sec})$ is compatible with a number of coherent
tubulins of order
\be
 N \simeq 10^{12}
\label{number}
\ee
provided that the
energy stored in the kink background is of the
order of
$eV$. This
is indeed
the case of the (dominant) sum of binding  and
resonant transfer energies $\Delta \simeq 1 eV $ (\ref{energy})
at room temperature
in the
phenomenological model of ref. \cite{mtmodel}.
This number of tubulin dimers
corresponds to a fraction of $10^{-7}$ of the total
brain,
which is pretty close to the fraction believed
to be responsible for human perception on the basis of
completely independent biological methods.
\pr
An independent estimate
for the
collapse time $t_{col}$,
can be given
on the basis of
point-like quantum gravity
quantum gravity theory, assuming that the latter exists, either
{\it per se}, or as an approximation to some string theory.
One incorporates quantum gravity effects by employing
wormholes \cite{coleman} in the structure of space time, and then
applies the calculus of ref. \cite{emohn} to infer the
estimates
of the collapse
time. In that case, one evaluates the off-diagonal elements
of the density matrix in real configuration space $x$,
which should be compared to that in string theory space
(\ref{asym}).
The result of ref. \cite{emohn} for the time of collapse
induced by the interaction with a `measuring apparatus'
with ${\cal N}$ units is
\be
  t_{coll}' \simeq \frac{1}{{\cal N}} (M_{gus}/m)^3
  \frac{1}{m^3 (\delta x)^2 }
\label{moh}
\ee
where $m$ is a typical mass unit in the problem.
The fundamental unit of velocities in (\ref{moh})
is provided by the velocity of light $c$, since
the formula (\ref{moh}) refers to generic
four-dimensional
space-time effects. In the case of the tubulin dimers,
it is reasonable to assume that the pertinent
moving mass is the effective mass $M^*$ (\ref{effmass})
of the kink background (\ref{eleven}). This
will make contact with the microscopic model above.
 To be specific, (\ref{effmass}) gives
 $M^* \simeq 3 m_p$,
where $m_p$ is the proton mass.
This makes plausible the rather daring  assumption
that the nucleons (protons, neutrons) themselves
inside the protein
dimers are the most sensitive constituents
to the effects of quantum gravity. This is a
reasonable assumption if one takes into account
that the nucleons are
much heavier than the (conformational) electrons.
If true,
this would really imply, then, that elementary particle scales
come into play in brain functioning.
In this picture, then, ${\cal N} $
is the number of tubulin dimers,
and moreover in our case $\delta x = O[4 nm]$, since the relevant
displacement length in the problem is of order of the
longitudinal dimension of each conformational pocket in the
tubulin dimer.
Thus,
substituting these
in (\ref{moh})
one derives the result  ${\cal N} = N \simeq 10^{12}$,
for the number of coherent dimers
that induce a collapse within ${\cal O}(1\,{\rm sec})$.
\pr
It is remarkable that the final numbers agree
between these two estimates. If one takes into account
the distant methods involved in the derivation of the
collapse times
in the respective approaches, then one realizes that
this agreement
cannot be a coincidence. Our belief is that
it reflects the fundamental r\^ole
of quantum gravity in the brain function.
\pr
At this stage,
it is important to make some clarifying remarks
concerning the kind of collapse that we advocate
in the framework of non-critical string theory.
In the usual model of collapse due to quantum gravity \cite{emohn},
one obtains an estimate of the collapse of the off-diagonal elements
of the density matrix in configuration space, but no information
is given for the diagonal elements.
In the string theory framework
of ref. \cite{emn} the collapse (\ref{asym})
also refers to off-diagonal elements of the string
density matrix, but in this case the configuration space
is the string background field space.
In this case the off-diagonal element collapse suffices, because
it implies {\it localization} in string background space,
which means that the quantum string chooses to settle down in
one of the classical backgrounds, which
in the case at hand is the solitonic background discussed
in section 2.
\pr
Of course,
the important question
`which specific background is
selected by the  above process ?' cannot be
answered unless a full string
field theory dynamics is developed. However, we believe that our
approach \cite{emn} of viewing the selection of a critical string
ground state as a generalized `measurement' process
in string theory space might prove advantageous over other dynamical
methods in this respect.

\section{Growth (Dynamical Instability)
of a Microtubule Network and Liouville Theory}
\pr
The above considerations are valid
for MT networks whose size of individual MT
is larger than a certain critical size \cite{mtmodel}.
Kink-like excitations, that were argued to be crucial
for the physics of the conscious functions of the brain,
cannot form for small microtubules.
The question, therefore, arises whether the non-critical
effctive string theory framework described above
is adequate for describing the growth process
associated with the formation of a MT network.
This phenomenon is physically  and biologically very
interesting since these structures are known
to be the only ones so far that exhibit
the so-called `{\it dynamical instability growth}' \cite{mitchison}.
This is an out-of-equilibrium process acording to which
an individual MT can switch randomly between an
`assembly state' (+), in which the MT grows with a
speed $v_+$,
and `diassembly state' (-), in which the MT shrinks
with velocity $v_{-}$.
Recently there have been attempts to construct
simple one-dimensional
theoretical models with diffusion~\cite{growth}
that can describe qualitatively the above phenomenon.
An interesting feature of these models, relevant to our framework,
is that for a certain range of their parameters exhibit
a phase transition to an unbounded growth
state\footnote{This situation may also be viewed from a
spontaneous-symmetry breaking point of view
along the lines of ref.
\cite{ssb}, which in one dimension can occur
when the system is out of equilibrium.
This second point of view is more relevant to
our non-critical string approach, which as we
explained earlier is an out-of-equilibrium
process.
The symmetry breaking can be exhibited
easily by looking at one-dimensional models with driven diffusion
of say two species of particles corresponding
to the (+) and (-) conformational
states of the MT growth.
In the broken phase there is a difference between the
`currents' correpsonding to the (+) or (-) states.
Under certain plaussible conditions \cite{ssb},
associated with formation of droplets of the
`wrong sign' in any given configuration of the
above states,
the system can
switch between (+) and (-) states
with a switching time that depends on
dynamical parameters. For a certain range of
these parameters the switching time is of order
$e^{N}$, where $N$ is the size of the system, thereby
implying spontaneous symmetry breaking
in the thermodynamic limit where $N \rightarrow  \infty$.}.
In the case of MT networks, it is known experimentally
that a `sawtooth' behaviour in the time-dependence
of the size of a MT (Fig. 3),
occurs as a result of hydrolysis
of GTP nulceotides bound to the tubulin proteins
(i.e. the transformation GTP  $\rightarrow$ GDP )\cite{mitchison}.
The hydrolysis is responsible for providing the
necessary free energy for the conformational changes of
the tubulin dimers: the dynamical instability phenomenon
pertains to polymerization of the GTP tubulin, while
the GDP tubulin stays essentially unpolymerized.
In view of quantum oscillations between the
two conformations of the tubulin, and the above
different behaviour of polymerization,
it is natural to conjecture that quantum
effects may play a r\^ole in the MT growth process.
\pr
Thus, our Liouville theory prepresentation
of the effective degrees of freedom $u$ involved
in the model of \cite{mtmodel}, invented to explain
classical aspects of the hydrolysis of GTP $\rightarrow$ GDP,
might, in principle, be able of explaining qualitatively
the `sawtooth' behaviour of Fig. 3, even before the formation
of kinks.
To this end, we remark
that in Liouville dynamics, with the Liouville
scale identified with the target time \cite{emn},
the inherent non-unitarity (in the world-sheet)
of the Liouville mode implies that
the central charge $Q$ of the theory flows
with the scale in such a way that
near fixed points it oscillates a bit
before settling down. Indeed,
for a non-critical string with running
central charge $C[g, t]$ , $t$ is the Liouville scale/time,
the following second order equation (local in target space-time)
holds near a fixed point of the Renormalization Group Flow~\cite{tseytl}
\be
 {\ddot C}[g, t]
 + Q [g,t] {\dot C} [g,t] \le 0 ~for C \ge 25
 \qquad ; \qquad  Q^2 [g, t] =\frac{1}{3}
(C[g,t ] - 25)
\label{ctheorem}
\ee
This is a local phenomenon in target space-time.
Globally,
there is a preferred direction
in time along which the
entropy of the system increases \cite{zam,emn}.
\pr
The small oscillations of $C$
in (\ref{ctheorem})
may be attributed to the double direction
of Liouville time that arises as a result
of imaginary parts (dynamical instabilities)
appearing due to
world-sheet
regularization by analytic continuation
of non-critical string correlation functions
\cite{kogan,emn} (Fig. 4)
This point is discussed briefly in Appendix A.
\pr
Along each direction of Liouville time in Fig. 4
there is an associated variation of $Q [g,t]$ and $C[g,t ]$,
and the volume of the `one-dimensional universe' of MT
either increases or decreases \cite{emninfl}. Thus,
as a result of the oscillations of the Liouville dynamics
(\ref{ctheorem}) a `sawtooth' behaviour of MT, which expresses the
result
of polymerization of GTP tubulin,
can be qualitatively
explained within the framework of non-critical Liouville dynamics.
The analogy of the
above situation
to a stochastic
expansion of the Universe (inflation)
in non-critical string theory as discussed in ref.
\cite{emninfl} should be pointed out.
\pr
It should be noted at this stage that
in this effective framework the origin of non-criticality
of the subsystem of tubulin dimers is left unspecified.
Quantum Gravity fluctuations appear on an equal footing
with the environment of the nucleating solvant that
sourrounds the MT in their physical environment.
The distinction can be made once a detailed
description
of the environment is given, which of course
would specify the form of the target metric
coupled to
the matter system of tubulins. As we have discussed in previous
sections,
in the quantum Gravity
case this is achieved by the
exactly marginal operators of the $SL(2,R)/U(1)$
conformal field theory that describes
space-time singularities in one-dimensional strings \cite{witt}.
The latter involve global (non-propagating)
string modes which cannot be detected
by localized scattering experiments \cite{emn}.
On the other hand, it seems likely that
an exact conformal field theory
that describes the nucleating solvant does not exist.
However, the effective Liouville string that describes
the embedding of a MT in it, and the associated
environmental entanglement, is obtained by a
simple Liouvlle dressing of the model discussed in section 3.
The information about the environment is hidden in the
form of the target metric that couples to the system.
The fact
that {\it both} the formation of an MT via the dynamical instability
phenomenon,
and the quantum
gravity effects on MT, involve conformational changes of the tubulin
supports the above point of view.
Of course the strength of the dynamical instability
in case the latter is due to quantum gravity fluctuations
will be much more suppressed as compared to the hydrolysis case.
In that case,
there will not be sufficient time
for complete polymerization
of the GTP tubulin conformation. In such a case
one would probably expect `sudden'
`sawtooth' peaks of the tip of the MT
whose magnitude will be affected by the order of
quantum gravity entanglement.
The growth in this case will be bounded.
\pr
It should be mentioned that recently there have been
some
experiments claiming such length fluctuations
\cite{kkk}, in the case of
carbon nanotubes. The authors of ref. \cite{kkk}
claimed that they observed such changes in the length
of the tubes, which they attributed to sudden jumps
of the respective wavefunctions, according to
the approach of Ghirardi Rimini and Weber \cite{grw}.
In this respect, one should think of repeating the tests
but with isolated MT\cite{rosu}.
We should stress, however, that the above discussion is at this stage
highly speculative and even controversial, given that
there appear to be conventional explanations
of this phenomenon in carbon nanotubes\cite{conv}.
Of course such conventional explanations
do not exclude the possibility of future observations of
quantum-gravity
induced bounded growth in MT,
along the lines
sketched above.
\pr
We close this section by mentioning that
in realistic situations the growth process of
an MT network is not unlimited.
After a critical length  is
exceeded, the growth is saturated and eventually stops.
The formation of kink excitations (\ref{eleven})
might be important for this
auto-regulation of the MT growth \cite{mtmodel}.
For more details on the conjectural r\^ole of the kinks
in this growth-control mechanism we refer the reader
to the literature \cite{mtmodel,growth}.
\pr
The above situation
should be compared with
the limitations on
the (stochastic) Universe expansion
in the non-critical-string-driven inflationary
scenario of ref. \cite{emninfl}.
There, it can be shown that within our framework
of identifying the Liouville field with the target time,
the average density
$<<\delta _s>> \equiv Tr(\rho \delta _s ) $
of the
non-critical strings
that drive the inflationary scenario
obey an equation of the form \cite{turok,emninfl}
\be
         \partial _t <<\delta _s >> = -aQ<<\delta _s >>
+ bQ^3 <<\delta _s >>
\label{denst}
\ee
where $\alpha $,$b$ are positive quantities,
computable in principle in
the Liouville-string framework \cite{emninfl}.
The first term in (\ref{denst})
is due to the (exponential) expansion
of the string-Universe volume, and the second term
corresponds to the regeneration of strings
via breaking of large strings whose size exceeds that of
the Hubble horizon \cite{emninfl}.
This second term comes from the diffusion due to the
non-quantum mechanical terms in the equation
(\ref{quantumliouv}).
At early times the diffusion term balances the
string depletion
effects
of the first term and the uniform density condition
for inflation (universe exponential expansion)
is satsfied. As the time elapses, however,
the depletion term in (\ref{denst})
dominates, the Universe's expansion is diminished gradually,
and eventually stops.
This is the case when the non-equilibrium non-critical string
approaches its (critical-string ) equilibrium state.
Hence, in our case one may
view (\ref{denst}) as an effective
model for the temporal evolution of the density of tubulin dimers.
Then, one can understand, at least qualitatively,
the above limitations of the MT growth process by the
presence of the kinks, since the latter can be associated with
an equilibrium ground state of the effective
string theory describing a MT.

\pr
\section{Conclusions}
\pr
We have presented an effective model for the
simulation of the dynamics of the tubulin dimers
in the brain. We have used an effective $(1 + 1)$-dimensional
string representation to
study the dynamics of
a detailed mechanism
for energy transfer in the
biological cells. We argued how it can give rise to
a large-scale coherent state in the dimer lattice.
Such a state
is obtained from quantization of
kink solitonic states that transfer energy
through the cell without dissipation.
The quantization became possible
through the freedom that string theory offers, enabling
one to
cast dynamical problems with friction in a
Hamiltonian form.
The collapse phenomenon in our approach does
not require the existence of a wave-fucntion,
and it
is induced by the formation of microscopic
black holes (singularities)
in the effective one-dimensional
space-time of the tubulin chains. This is achieved
by the dynamical collapse of pulses of the
displacenent field of the MT dimers. The pulses are a result of
abrupt conformational changes ($\alpha\leftrightarrow\beta$)
that sufficiently
distort the surrounding space-time.
In this sense, the situation is similar but not identical, to
the `sudden hits' that a particle's wave function suffers
occasionally (every $10^8$ years)
in the model
of quantum measurement
of Girardi, Rimini and Weber \cite{grw}.
However, contrary to these conventional theories,
our stringy approach to gravity-induced collapse
\cite{emn}
incorporates automatically
an irreversible flow of time
for specifically {\it stringy} reasons, and {\it
energy conservation}.
When applied to the model of MT, our approach
implies a collapse time of
${\cal O}(1\,{\rm sec})$,
which is obtained by the interaction of a tubulin dimer with
a fraction of $10^{-7}$
of the total number of tubulin dimers in the brain.
This number is fairly close to the fraction of the
brain that neuroscientists believe responsible
for human perception.
This is a very strong indication
that the above ideas, although speculative at this stage,
might be relevant for the discovery
of a physical model for consciousness and its relation
to the irreversible flow of time.
In addition, our model predicts
damped
(microscopic)
quantum-gravity-induced oscillations
of the length of isolated MTs, which
are due to the different properties
of the two tubulin conformations
under polymerization (phenomenon
of bounded dynamical-instability growth).
\pr
There are certain formal aspects of our effective model,
namely its two-dimensional structure and its complete
(quantum) integrability, that might turn out to be
important
features for the construction of realistic soluble
models for
brain function. The quantum
integrability is due to generalized infinite-dimensional
symmetry structures ($W$-symmetries and its generalizations)
which are strongly linked with issues
of quantum coherence and unitary evolution in phase-space.
Such symmetries are related to global non-propagating modes
of the effective string theory, which do not decouple
from the propagating (observed) modes in the presence of
(microscopic)
space-time singularities.
It  will be interesting to
understand further the physical r\^ole of such structures
in the models of MT. At present they appear as an environment
of fundamental string modes that are physical at Planck scales.
However, such structures, may admit a less-ambitious
physical meaning, associated with fundamental biological
structures
of the
brain.
We should stress that
the entire picture of non-critical
string we have described above, which is a
model-independent picture as far as environmental
operators are concerned, could still apply in such cases,
but
simply describing
purely biological
environmental entanglement of the conscious part
of the brain, the latter being described as a
completely integrable model.
We do hope to come back to these issues
in the near future.
\pr
\newpage
\nk {\Large{\bf  Acknowledgements}}
\pr
It is a pleasure to acknowledge discussions
with J. Ellis.
We also acknowledge useful discussions and
correspondence with
S. Hameroff. N.E.M. acknowledges informative communications
with H. Rosu on quantum jumps and experimental observations of
dynamical
instabilities in nanotubes.
The work of N.E.M. is supported by a EC Research Fellowship,
Proposal Nr. ERB4001GT922259 .
That of D.V. N. is partially supported by D.O.E. Grant
DEFG05-91-GR-40633.
\pr
\newpage

\nk {\Large {\bf Appendix A} }
\pr
\nk {\bf Extracts from
Non-Critical (Liouville) String Theory  and  Time
as the Liouville scale}
\pr
In this appendix we shall comment on the non-factorizability
of the induced target-space $\nd{S}$-matrix for matter
scattering in a non-conformal string background.
This reflects information leakage as a result of the
non-critical character of the string. Although  for our purposes
primarily we shall be interested in a specific string background,
that of a stringy black hole, however in this section
our discussion will be kept as general as possible with the
aim of demonstrating the generality of our scheme.
\pr
Consider a conformal field theory on a two-dimensional
world sheet, described by an action $S[g^*]$.
The $\{ g^* \}$  are a set of space-time backgrounds.
The theory is perturbed by a deformation $V_i$,
which is not conformal invariant
\be
   S[g] = S[g^*] + \int d^2 z g^i V_i
\label{C1}
\ee
The couplings $g^i$ corrspond to world-sheet
renormalization group $\beta$-functions
\be
   \beta ^i = (h_i - 2)
   (g^i - (g^{*})^i) +
c^i_{jk}(g^j - (g^{*})^j) (g^k - (g^{*})^k) +  \dots
\label{C2}
\ee
expressing the scale dependence of the non-conformal
deformations.
The operator product expansion coefficients
are defined as usual  by coincident limits in the product
of two vertex operators $V_i$
\be
lim_{\sigma \rightarrow 0} V_j (\sigma )V_k (0) \simeq
c^i_{jk} V_i (\frac{\sigma}{2}) + \dots
\label{C3}
\ee
where the completeness of the set $\{ V_i \}$ is assumed.
\pr
Coupling the theory (\ref{C1}) to two-dimensional quantum gravity
restores the conformal invariance at a quantum level,
by making the gravitationally-dressed operators
$[V_i]_{\phi}$ {\it exactly} marginal, i.e. ensuring
the absence of any covariant
scale dependence with respect to the
world-sheet metric $\gamma _{\alpha\beta}$.
Below we simply outline the basic results,
used in our approach here.

One rescales the world-sheet metric
\be
    \gamma _{\alpha\beta} = e^{\phi} {\widehat \gamma}_{\alpha\beta}
\label{C4}
\ee
with ${\widehat \gamma}$ is kept fixed,
and then one integrates over the Liouville mode $\phi $.
The measure of such an integration\cite{DDK}
can be expressed in terms of the fiducial metric
${\widehat \gamma }$ by means of a determinant
which is the exponential of the Liouville action.
The final result for the gravitationally-dressed
matter theory is then
\be
S_{L-m} = S[g^*] + \frac{1}{4\pi\alpha '}
\int d^2 z \partial _\alpha \phi \partial ^\alpha \phi
- QR^{(2)} + \lambda ^i(\phi ) V_i
\label{C5}
\ee
where $\alpha '$ is the Regge slope for the world-sheet theory
(inverse of the string tension).
The gravitational dressing of the operators follows
from the requirement of restoring
the conformal invariance of the theory, at
any given order in the coupling-constant
expansion. For instance, to order $O(g^2)$
the gravitationally-dressed
coupling $\lambda (\phi )$ are given by\cite{DDK,schmid}:
\be
\lambda ^i(\phi ) =g^i e^{\alpha _i \phi }
+ \frac{\pi}{Q \pm 2\alpha _i } c^i_{jk} g^jg^k
\phi e^{\alpha _i \phi } + \dots
\label{C6}
\ee
with
\be
Q=\sqrt{\frac{|25-c|}{3}}  \qquad ; \qquad
\alpha _i ^2 + \alpha _i Q =sqn(25-c)(h_i - 2)
\label{C7}
\ee
and $c$ is the (constant) central charge of the
non-critical string.
 From the quadratic equation for $\alpha _i$
only the solution
\be
\alpha _i = -\frac{Q}{2} +
\sqrt{\frac{Q^2}{4} - (h_i - 2)}
\label{solutions}
\ee
for $c \ge 25 $, is kept due to
the Liouville
boundary conditions.

In ref. \cite{emn} we made an extra assumption, as compared
to the above standard Liouville dynamics. We identified
the field $\phi $ with a dynamical local scale on the
world sheet. This induces extra counterterms
in the world-sheet renormalized action. Consistency
of the scheme required that the Liouville $\beta $
functions are identical with the flat space renormalization
coefficients upon the replacement $g^i \rightarrow \lambda (\phi)^i$.

The type of operators that we are interested in this
work, are such that $h_i =2$ but $c^i_{jk} \ne 0$.
In the language of conformal field theory
this means that these operators are $(1, 1)$
but {\it not exactly marginal}.
 From (\ref{C6}), then, one obtains the simple relation
\be
   \frac{d \lambda ^i(\phi) }{d t_p} =  \beta ^i
\label{C8}
\ee
where the time $t_p$ is related to the Liouville mode $\phi $
as
\be
        t_p = -\frac{1}{\alpha Q}{\rm ln}A \qquad ; \qquad
  A \equiv \int d^2z \sqrt{{\hat \gamma}}
e^{\alpha \phi (z,{\bar z})}
\qquad ; \qquad \alpha = -\frac{Q}{2} + \frac{1}{2}\sqrt{Q^2 + 8}
\label{tphys}
\ee
with $A$ the world-sheet area.
In the local scale formalism of ref. \cite{emn}
$Q$ is given by
\be
Q=\sqrt{\frac{|25-C[g,\phi]|}{3} + \frac{1}{2}\beta ^i G_{ij} \beta ^j}
\label{C9}
\ee
where $C[g,\phi ]$ is the Zamolodchikov $C$-function
\cite{zam}, which reduces to the
central charge $c$ at a fixed point of the flow.
The extra terms in (\ref{C9}), as compared to (\ref{C7}), are due to
the local character of the renormalization group scale\cite{emn}.
Such terms may always be removed by non-standard redefinitions
of $C[g,\phi ]$. The quantity
$G_{ij}$ is related to divergencies
of the two-point functions $<V_iV_j>$
and hence to Zamolodchikov metric
\cite{zam,emn}.

 From the renormalization-group structure
(\ref{C6}) one obtains close to a fixed point \cite{tseytl}
\be
  {\ddot \lambda (\phi )}^i + Q {\dot \lambda} ^i = -\beta ^i = - G^{ij}
  \partial _i C [\lambda, \phi]
\label{C10}
\ee
where the dot denotes differentiation
with repsect to the Liouville local scale $\phi $.

For the $C[g,\phi ]$  (local in target space-time)
one obtains near a fixed point
\be
 {\ddot C}[g, t]
 + Q [g,t] {\dot C} [g,t] \le 0 ~for C \ge 25
 \qquad ; \qquad  Q^2 [g, t] =\frac{1}{3}
(C[g,t ] - 25)
\label{C11}
\ee
The small oscillations of $C[\lambda, \phi ]$,
before it settels down to a fixed point,
are due to the `non-unitary' world-sheet contributions
of the Liouville mode $\phi$ ; however
globally in target space-time
there is a monotonic change of the degrees of freedom of the
system, as discussed in detail in \cite{emn}.

These considerations can be understood more easily
if one looks at the correlation functions
in the Liouviulle theory, viewing the liouville field
as a local scale on the world sheet .
Standard computations\cite{goulian} yield for an $N$-point correlation
function among world-sheet integrated
vertex operators $V_i\equiv \int d^2z V_i (z,{\bar z}) $ :
\be
A_N \equiv <V_{i_1} \dots V_{i_N} >_\mu = \Gamma (-s) \mu ^s
<(\int d^2z \sqrt{{\hat \gamma }}e^{\alpha \phi })^s {\tilde
V}_{i_1} \dots {\tilde V}_{i_N} >_{\mu =0}
\label{C12}
\ee
where the tilde denotes removal of the
Liouville  field $\phi $ zero mode, which has been
path-integrated out in (\ref{C12}).
The world-sheet scale $\mu$ is associated with cosmological
constant terms on the world sheet, which are characteristic
of the Liouville theory \cite{DDK}.
The quantity $s$ is the sum of the Liouville anomalous dimensions
of the operators $V_i$
\be
s=-\sum _{i=1}^{N} \frac{\alpha _i}{\alpha } - \frac{Q}{\alpha}
\qquad ; \qquad \alpha = -\frac{Q}{2} + \frac{1}{2}\sqrt{Q^2 + 8}
\label{C13}
\ee
The $\Gamma $ function can be regularized\cite{kogan,emn}
(for negative-integer
values of its argument) by
analytic coninuation to the complex-area plane using the
the Saaschultz contour
of Fig. 4. This yields the possibility
of an increase of the running central charge
due to the induced oscillations of the dynamical
world sheet area (related to the Liouville zero mode).
This is associated with the oscillatory solution
(\ref{C11}) for the Liouville central charge.
On the other hand, the bounce intepretation
of the infrared fixed points of the flow,
given in refs. \cite{kogan,emn},
provides an alternative picture
of the overall monotonic change
at a global level in target space-time.

The above formalism also allows for an
explicit demonstration of the
non-factorizability of the
superscattering matrix associated with
target-space interactions in non-critical
string theory. This
was very important for our purposes
in the context of the collapse
of the wave-function as a result of quantum
entanglement due to quantum gravity fluctuations.

To this end, one expands the Liouville
field in (normalized) eigenfunctions  $\{ \phi _n \}$
of the Laplacian $\Delta $ on the world sheet
\be
 \phi (z, {\bar z}) = \sum _{n} c_n \phi _n  = c_0 \phi _0
 + \sum _{n \ne 0} \phi _n \qquad \phi _0 \propto A^{-\frac{1}{2}}
\label{C14}
\ee
with $A$ the world-sheet area,
and
\be
   \Delta \phi _n = -\epsilon_n \phi _n  \qquad n=0, 1,2, \dots
\qquad (\phi _n, \phi _m ) = \delta _{nm}
\label{C15}
\ee
The result for the correlation functions (without the Liouville
zero mode) appearing on the right-hand-side of eq. (\ref{C12})
is, then
\bea
{\tilde A}_N \propto &\int & \Pi _{n\ne0}dc_n exp(-\frac{1}{8\pi}
\sum _{n\ne 0} \epsilon _n c_n^2 - \frac{Q}{8\pi}
\sum _{n\ne 0} R_n c_n + \nn \\
{}~&~&\sum _{n\ne 0}\alpha _i \phi _n (z_i) c_n )(\int d^2\xi
\sqrt{{\hat \gamma }}e^{\alpha\sum _{n\ne 0}\phi _n c_n } )^s
\label{C16}
\eea
with $R_n = \int d^2\xi R^{(2)}(\xi )\phi _n $. We can compute
(\ref{C16}) if we analytically continue \cite{goulian}
$s$ to a positive integer $s \rightarrow n \in {\bf Z}^{+} $.
Denoting
\be
f(x,y) \equiv  \sum _{n,m~\ne 0} \frac{\phi _n (x) \phi _m (y)}
{\epsilon _n}
\label{fxy}
\ee
one observes that, as a result
of the lack of the zero mode,
\be
   \Delta f (x,y) = -4\pi \delta ^{(2)} (x,y) - \frac{1}{A}
\label{C17}
\ee
We may choose
the gauge condition  $\int d^2 \xi \sqrt{{\hat \gamma}}
{\tilde \phi }=0 $. This determines the conformal
properties of the function $f$ as well as its
`renormalized' local limit\cite{lag}
\be
   f_R (x,x)=lim_{x\rightarrow y } (f(x,y) + {\rm ln}d^2(x,y))
\label{C18}
\ee
where  $d^2(x,y)$ is the geodesic distance on the world sheet.
Integrating over $c_n$ one obtains
\bea
{}~&& {\tilde A}_{n + N} \propto
exp[\frac{1}{2} \sum _{i,j} \alpha _i \alpha _j
f(z_i,z_j) + \nn  \\
{}~&&\frac{Q^2}{128\pi^2}
\int \int  R(x)R(y)f(x,y) - \sum _{i} \frac{Q}{8\pi}
\alpha _i \int \sqrt{{\hat \gamma}} R(x) f(x,z_i) ]
\label{C19}
\eea

We now consider
infinitesimal Weyl shifts of the world-sheet metric,
$\gamma (x,y) \rightarrow \gamma (x,y) ( 1 - \sigma (x, y))$,
with $x,y$ denoting world-sheet coordinates.
Under these,
the correlator $A_N$
transforms as follows\cite{emnest}
\bea
&~&
\delta {\tilde A}_N \propto
[\sum _i h_i \sigma (z_i ) + \frac{Q^2}{16 \pi }
\int dx \sqrt{{\hat \gamma }} {\hat R} \sigma (x) +    \nn \\
&~&
\frac{1}{{\hat A}} \{
Qs \int dx \sqrt{{\hat \gamma }} \sigma (x)
       +
(s)^2 \int dx \sqrt{{\hat \gamma }} \sigma (x) {\hat f}_R (x,x)
+  \nn \\
&~&
Qs \int \int d^2x d^2y
\sqrt{{\hat \gamma }} R (x) \sigma (y) {\hat {\cal
 G}} (x,y) -
  s \sum _i \alpha _i
  \int d^2x
  \sqrt{{\hat \gamma }} \sigma (x) {\hat {\cal
 G}} (x, z_i) -   \nn \\
&~&
 \frac{1}{2} s \sum _i \alpha _i{\hat f}_R (z_i, z_i )
  \int d^2x \sqrt{{\hat \gamma }} \sigma (x)
-    \nn \\
&~&
 \frac{Qs}{16\pi} \int
  \int d^2x d^2y \sqrt{{\hat \gamma (x)}{\hat \gamma }(y)}
  {\hat R}(x) {\hat f}_R (x,x) \sigma (y)\} ] {\tilde A }_N
\label{dollar}
\eea
where the hat notation denotes transformed quantities,
and
the function  ${\cal G}$(x,y)
is defined as
\be
  {\cal G}(z,\omega ) \equiv
f(z,\omega ) -\frac{1}{2} (f_R (z,z) + f_R (\omega, \omega ) )
\label{C20}
\ee
and transforms simply under Weyl shifts\cite{lag}.
We observe from (\ref{dollar}) that
if the sum of the anomalous dimensions
$s \ne 0$ (`off-shell' effect of
non-critical strings), then there are
non-covariant terms in
(\ref{dollar}), inversely proportional to the
finite-size world-sheet area $A$.
In general, this is a feature of non-critical strings
wherever the Liouville mode is viewed as a local scale
of the world sheet. In such a case,
the central charge of the theory
flows continuously with time/scale $t$,
as a result of the Zamolodchikov
$c$-theorem \cite{zam}. In contrast, the screening operators
yield quantized values\cite{aben}.
This induced  time ($A$-) dependence
of the correlation function $A_N$ implies the
breakdown of their interpretation as
factorisable $\nd{S}$-matrix elements.
\pr
In our framework, the effects of the quantum-gravity
entanglement induce such $A$-dependences
in correlation functions of the propagating
matter vertex operators of the string \cite{emn},
correspondng to the displacement field $u(x,t)$ of the MTs.
To this end, we first note that
the physical states in such completely integrable
models fall into representations
of the $SL(2,R)$ target symmetry, which are classified by
the non-compact isospin $j$ and its third component
$m$\cite{distler}. There is a formal
equivalence of the physical states
between the flat-space time $(1 + 1)$-dimensional string
and the black-hole model\footnote{Originally, there were claims
\cite{distler}
that there are extra states
in the black hole models, as compared to the flat-space time
string; however, later on it has been shown that
such states
can be either
gauged away\cite{eguchi} or
boosted\cite{ardalan},
and so they disappear from the physical spectrum.},
which confirms the point of view\cite{emnsel,emn}
that the flat-space $(1+1)$-dimensional string
theory is the spatially- (and temporally-) asymptotic limit of the
$SL(2,R)/U(1)$ black hole.
The existence of discrete (quasi-topological, non-propagating)
Planckian modes in the two-dimensional string theory
leads to {\it selection} rules\cite{emnsel}
in the number $N$ of the scattered propagating degrees of freedom,
according to the intermediate-exchange state :
\bea
{}~&~&{\rm Asymptotic~correspondence}~(\epsilon _\phi~(p) \equiv {\rm
Liouville~energy~(momentum)}): \nn \\
{}~&~&j \rightarrow \epsilon_\phi ,~m \rightarrow
  \frac{3 p}{2\sqrt{2}}
\nn \\
{}~&~&{\rm Asymptotic~Kinematics}:
\nn \\
{}~&~&\sum _{i=1}^{N-1} p_i=\frac{N-2}{\sqrt{2}} \qquad ; \qquad
p_N =-\frac{N-2}{\sqrt{2}} \qquad; \qquad N \ge 3
\nn \\
{}~&~&{\rm Selection~Rules}: \nn \\
{}~&~&j=\frac{1}{3}m - 1 + \frac{1}{2} (N -2)~,~j \ge \frac{1}{4}(N- 5)
\label{sel}
\eea
Such rules are obtained by imposing the Liouville
energy $\epsilon _\phi $ and momentum $p$
conservation, leading to $s =0$, with $s$ the sum of
Liouville anomalous dimensions as defined earlier.
Obviously, if the exchange state is an (off-shell)
propagating
mode, belonging to the continuous representation
of $SL(2,R)$, i.e. $j \in {\bf R}$, $j \ge -\frac{1}{2}$, there
are no restrictions
on $N$, and a convetnional $S$-matrix amplitude
can be defined as the residue of the Liouville
amplitudes with respect to the single poles in $s$~\cite{pol}.
However, in two space-time dimensions graviton excitations
are {\it discrete}, corresponding to string-level-one
representations  of $SL(2,R)$.
Hence,  once non-trivial quantum-gravity  fluctuations
are considered in our approach, which in two dimensions
are black-hole backgrounds (\ref{st11}),
one has to take into account  discrete
on-shell exchange modes in
the Liouville correlation functions\footnote{The on-shell condition
imposes algebraic relations among $j$ and $m$ for such modes,
involving the string-level number due to the Virasoro constraints
\cite{distler}. This, in turn, implies restrictions
to the number $N$
of scattered particles in such cases.}. Such states represent
{\it excited} states of the (virtual) black-holes, created
by the collapse of the propagating matter modes $u(x,t)$,
as described in section 4 (\ref{st11}).  In two-dimensional
string theory black holes are like particles\cite{emn}, the
difference being their topological nature.
Such modes constitute, in our case, `the consciousness
degrees of freedom', which cannot be measured
by local scattering experiments. Integrating them out
in the `mind', implies a time arrow as described in ref.
\cite{emn}. Indeed, in the correlation functions
(\ref{C12}),
as a result of Liouville energy conservation
(\ref{sel}), one of the modes is necessarily discrete.
If we suppress such modes, and consider only
external propagating modes, accessible to
physical scattering processes,
then it is evident
that $s \ne 0$. According to our previous
analysis (\ref{dollar}), then, this
implies world-sheet-area($A$) dependence
of the correlation functions.
In this picture,
we also note that
Quantum-Gravitational fluctuations of singular space-time
form, correspoding to higher-genus world-sheet effects,
have been argued~\cite{emn} to be represented
{\it collectively}
by world-sheet instanton-anti-instanton deformations
in the stringy $\sigma$-model.
It is known \cite{yung} that such configurations
are responsible for a {\it non-perturbative} breakdown
of the conformal invariance of the $\sigma$-model.
Using a dynamical (world-sheet) renormalization-group
scale (Liouville mode) to represent
all
such non-conformal invariant
effects\cite{emn}, and identifying it
with the target time,
one, then, arrives
at non-factorisable superscattering
operators, as described above.
\pr
Notice that the precise microscopic
nature of the environmental operators is not
essential as long as the latter imply a conformal
anomaly. There are general consequences of this
conformal anomaly,
including dynamical collapse of the string theory
space to a certain configuration, as discused in
section 4. A similar situation occurs in
ordinary
quantum mechanics of open systems.
Once a stochastic framework
using state vectors
is adopted \cite{gisin} for the
description of environmental effects,
there will always be localization
of the state vector in one of its
channels, irrespective of the detailed form of the
environment operators.
It should be noted that stochasticity
is a crucial feature of our approach
too.
This follows from the stochastic nature of
the renormalization group in two-dimensions
\cite{friedan,emn}.
This stochasticity was argued to play an important r\^ole
in the MT growth, discussed in section 5.

\newpage
\pr
\nk {\Large {\bf Appendix B} }
\pr
\nk {\bf  Variational Approach to Soliton Quantization
via Squeezed Coherent States }
\pr
It is the purpose of this
appendix to discuss briefly the formalism
leading to the quantization of the solitonic
states discussed in section 2.
\pr
One assumes the existence of a canonical
second quantized formalism for the $(1+1)$-dimensional
scalar field $u(x,t)$, based on creation and
annihilation operators
$a^\dagger _k$,
$a_k$.
One then constructs a squeezed
vacuum state\cite{tdva}
\be
     |\Psi (t) > = N(t) e^{T(t)} |0>
\qquad ; \qquad T(t) = \frac{1}{2} \int \int
dx dy u(x) \Omega (x,y,t) u(y)
\label{sq1}
\ee
where $|0>$ is the ordinary vacuum state
annihilated by $a_k$,
and
$N(t)$ is a normalization factor to be determined.
$\Omega (x,y,t)$ is a complex function,
which can be splitted
in real and imaginary parts
as
\bea
  \Omega (x,y,t) &=& \frac{1}{2}[G_0^{-1}(x,y) - G^{-1}(x,y,t)]
+ 2i\Pi (x,y,t) \nn \\
G_0 (x,y) &=& <0|u(x)u(y)|0>
\label{sq2}
\eea
The squeezed coherent state for this system can be then defined
as\cite{tdva}
\be
 |\Phi (t) > \equiv e^{iS(t)}|\Psi (t)> \qquad ; \qquad
S(t) = \int_{-\infty}^{+\infty} dx[D(x,t)u(x) - C(x,t)\pi (x)]
\label{sq3}
\ee
with $\pi (x)$ the momentum conjugate to $u(x)$, and
$D(x,t)$, $C(x,t)$ real functions.
With respect to this state $\Pi (x,t)$ can
be considered as a momentum canonically conjugate
to $G(x,y,t)$ in the following sense
\be
         <\Phi (t) | -i\frac{\delta}{\delta \Pi (x,y,t)}
|\Phi (t)> = - G(x,y,t)
\label{sq4}
\ee
The quantity $G(x,y,t)$ represents the modified
boson field around the soliton.
\pr
To determine $C$,$D$, and $\Omega$ one applies
the Time-Dependent Variational Approach
(TDVA)~\cite{tdva} according to which
\be
\delta \int _{t_1}^{t_2} dt <\Phi (t) |(
i\partial _t - H) |\Phi (t) >  = 0
\label{sq5}
\ee
where $H$ is the canonical Hamiltonian of the system.
This leads to a canonical set of  (quantum) Hamilton
equations
\bea
{\dot D}(x,t)&=&-\frac{\delta {\cal H}}{\delta
C(x,t)} \qquad  {\dot C}(x,t) = \frac{\delta {\cal H}}
{\delta D (x,t)} \nn \\
{\dot G}(x,y,t) &=& \frac{\delta {\cal H}}{\delta \Pi (x,y,t)}
\qquad {\dot \Pi}(x,y,t) =\frac{\delta {\cal H}}{\delta
G(x,y,t)}
\label{sq6}
\eea
where the quantum energy functional
${\cal H}$ is given by\cite{tdva}
\be
  {\cal H} \equiv <\Phi (t) | H | \Phi (t) >= \int _{-\infty}
^{\infty} dx {\cal E} (x)
\label{sq7}
\ee
with
\bea
{\cal E} (x) = \frac{1}{2} D^2 (x,t) &+& \frac{1}{2} (
\partial _x  C(x,t))^2  + {\cal M}^{(0)} [C(x,t)] + \nn \\
\frac{1}{8}<x|G^{-1}(t)|y> &+& 2 <x|\Pi (t) G(t) \Pi (t) |y>
+\frac{1}{2}lim_{x \rightarrow y} \nabla _x\nabla _y <x| G(t) |y>
 - \\
 \frac{1}{8}<x|G_0^{-1}|y> &-& \frac{1}{2}lim _{x \rightarrow y}
\nabla _x \nabla _y <x|G_0 (t)|y>
\label{sq8}
\eea
where we use the following operator notation in coordinate
representation
$A (x,y,t) \equiv <x|A(t)|y>$, and
\be
M^{(n)} = e^{\frac{1}{2}(G(x,x,t)-G_0(x,x))\frac{\partial ^2}
{\partial z^2}} U^{(n)}(z) |_{z=C(x,t)}
\qquad ; \qquad U^{(n)} \equiv d^n U/d z^n
\label{sq9}
\ee
Above, $U $ denotes the potential of the original soliton
Hamiltonian, $H$.
Notice that the quantum energy functional is conserved
in time, despite the various time dependences
of the quantum fluctuations. This is a consequence
of the canonical form (\ref{sq6}) of the Hamilton equations.
\pr
Performing the functional derivations in (\ref{sq6})
one can get
\bea
 {\dot D}(x,t) &=& \frac{\partial ^2}{\partial x^2}
C(x,t) - {\cal M}^{(1)}[C(x,t)]   \nn \\
{\dot C}(x,t) &=& D(x,t)
\label{sq10}
\eea
which after elimination of $D(x,t)$, yields
the modified (quantum) soliton equation (\ref{22c}).
We note that the quantities ${\cal M}^{(n)}$ carry
information about the quantum corrections, and in
this sense tha above scheme is more accurate
than the $WKB$ approximation \cite{WKB}.
The whole scheme may be thought of as
a mean-field-approach to quantum corrections to the soliton
solutions.
\pr
In our string
framework, then,
these point-like quantum solitons can be viewed
as a low-energy approximation to some more general
ground state solutions of a
non-critical string theory, formulated in higher genera on the
world sheet to account for the quantum corrections.
We have not worked out in this work the full string-theory
representation of the relevant quantum coherent state.
This will be an interesting topic to be studied in the future,
which will allow for a rigorous
study of the effects of
the global string modes on the collapse of
the quantum-coherent preconscious state.

\newpage
{\Large {\bf Figure Captions}}
\pr

\nk {\Large {\bf Figure 1 }}- Microtubular Arrangement :
(a) the structure of a Microtubule (MT), (b)
cross section of a MT, (c) two neighboring dimers  along
the direction of a MT axis
\pr

\nk {\Large {\bf Figure 2 }}- The two conformations
$\alpha $ and $\beta$ of a MT dimer.
Transition (switching)
between these two conformational states
can be viewed as a quantum-mechanical effect.
Quantum-Gravity entanglement can cause the collapse
of quantum-coherent states of such conformations,
which might arise
in a MT network modelling the preconscious state of
the human brain.

\pr
\nk {\Large {\bf Figure 3 }}- Illustration of the phenomenon
of `dynamical instability' of a MT network : (a) unbounded
`sawtooth' growth  (b) bounded `sawtooth' growth.
Dotted lines show the average over many MT with the
same dynamical parameters.

\pr
\nk {\Large {\bf Figure 4 }}- (a) Contour
of integration in the analytically-continued
(regularized) version of $\Gamma (-s)$ for $ s \in Z^+$.
This is known in the literature as the Saalschutz contour,
and has been used in
conventional quantum field theory to relate dimensional
regularization to the Bogoliubov-Parasiuk-Hepp-Zimmermann
renormalization method,
(b) schematic repesentation
of the evolution of the world-sheet area as the renormalization
group scale moves along the contour of fig. 4(a)

\end{document}